\documentclass[sigconf]{acmart}

\AtBeginDocument{%
  \providecommand\BibTeX{{%
    \normalfont B\kern-0.5em{\scshape i\kern-0.25em b}\kern-0.8em\TeX}}}

\begin{document}

\title{Matching Representations of Explainable Artificial Intelligence and Eye Gaze for Human-Machine Interaction}

\author{Tiffany Hwu}
\email{tjhwu@hrl.com}
\affiliation{%
  \institution{HRL Laboratories, LLC}
  \city{Malibu}
  \state{CA}
  \country{USA}
}

\author{Mia Levy}
\email{mmlevy@hrl.com}
\affiliation{%
  \institution{HRL Laboratories, LLC}
  \city{Malibu}
  \state{CA}
  \country{USA}
}

\author{Steven Skorheim}
\email{swskorheim@hrl.com}
\affiliation{
  \institution{HRL Laboratories, LLC}
  \city{Malibu}
  \state{CA}
  \country{USA}
}

\author{David Huber}
\email{djhuber@hrl.com}
\affiliation{%
  \institution{HRL Laboratories, LLC}
  \city{Malibu}
  \state{CA}
  \country{USA}
}


\begin{abstract}
Rapid non-verbal communication of task-based stimuli is a challenge in human-machine teaming, particularly in closed-loop interactions such as driving. To achieve this, we must understand the representations of information for both the human and machine, and determine a basis for bridging these representations. Techniques of explainable artificial intelligence (XAI) such as layer-wise relevance propagation (LRP) provide visual heatmap explanations for high-dimensional machine learning techniques such as deep neural networks. On the side of human cognition, visual attention is driven by the bottom-up and top-down processing of sensory input related to the current task. Since both XAI and human cognition should focus on task-related stimuli, there may be overlaps between their representations of visual attention, potentially providing a means of nonverbal communication between the human and machine. In this work, we examine the correlations between LRP heatmap explanations of a neural network trained to predict driving behavior and eye gaze heatmaps of human drivers. The analysis is used to determine the feasibility of using such a technique for enhancing driving performance. We find that LRP heatmaps show increasing levels of similarity with eye gaze according to the task specificity of the neural network. We then propose how these findings may assist humans by visually directing attention towards relevant areas. To our knowledge, our work provides the first known analysis of LRP and eye gaze for driving tasks.
\end{abstract}



\maketitle

\section{Introduction}
Human cognition is flexible and adaptive, learning from past experiences and generalizing to novel situations. Machines, on the other hand, can tirelessly process lifetimes of data for pattern recognition. The complementary strengths of humans and machines makes for ideal teamwork, but several obstacles exist in connecting the two disparate systems of information processing. A prerequisite for teaming in such situations is an examination of the mental representations of humans and machines to find common ground for communication. When performing certain tasks, the inputs that the human and machine find salient often overlap. Areas of machine salience can be found by explainable AI techniques that explain the inner workings of machine learning algorithms, while areas of salience in human cognition can be found through physiological monitoring techniques such as gaze tracking. The work described in this paper performs such comparisons on a dataset of driver eye gaze, revealing insightful patterns that expand possible applications in adaptive teaming.

\section{Background}
\label{background}

\subsection{Explainable AI (XAI)}
The term ``explainable artificial intelligence" (XAI) refers to methods that explain the processes used by artificial intelligence and machine learning and the results that they produce \citep{gunning2017explainable}. Particularly for high-dimensional machine learning methods such as deep learning, the methods are a black box, which often leads to mistrust and concerns in safety-critical applications, such as driving. XAI can improve the safety and trustworthiness of AI by showing the user or the developer why the AI is making certain decisions, which allows humans to validate and improve the AI's decisions. One specific method of XAI is known as layer-wise relevance propagation (LRP) \citep{bach2015pixel}, which can be applied to any multi-layered classifier, decomposing an output prediction into the contributions of each component of each layer. In the case of image classification, LRP displays a heatmap of which parts of an image contributed the most to the classification.

\subsection{Eye Tracking}
Eye tracking is a way of approximating human attention by measuring the focal point of human gaze as the human views an image or performs a task. It is most often done with hardware that detects pupil position using the infrared or visible light spectrum and aligning it with gaze points during a calibration period, after which human gaze can be estimated. The visual features that drive gaze can be divided into sources of bottom-up and top-down categories \citep{borji2012boosting}. Bottom-up features are visually salient and may have high contrast, brightness, or motion/flicker, whereas top-down features are related to the task at hand and are driven by the human's mental models and goals. In addition, top-down features are found when the human is attentive to the task at hand, whereas bottom-up features are more common in inattentive situations. In the case of driving, both types of features are encountered, including unexpected pedestrians and traffic signals.
Similarly, most eye tracking datasets have varying levels of bottom-up and top-down influences, where some are more task-specific and others are recorded on free viewing. For instance, the SALICON dataset is a free viewing dataset collected from the mouse movements of online subjects \citep{jiang2015salicon}. There are also other free viewing datasets based on actual human gaze, such as the CAT2000 dataset \citep{borji2015cat2000}, in which subjects view a collection of indoor and outdoor scenes, line drawings, and low resolution images. For predicting gaze in such free viewing tasks, the DeepGaze II gaze estimation model extracts high-level features found from neural networks trained on image recognition \citep{Kummerer_2017_ICCV}. The DeepGaze II model shows the potential of extracting information from machine learning methods to aid in human state estimation in the domain of free viewing.

The DR(eye)VE dataset includes eye tracking data recorded from several human subjects driving a car in naturalistic environments of varied settings, including different times of the day, settings, and weather conditions \citep{alletto2016dr,palazzi2018predicting}. It differs from the previously discussed datasets as it is task-specific. Instead of viewing images passively, the subject views a real scene while performing a relevant task. This work uses the DR(eye)VE dataset, which contains over 500,000 frames of data captured from eye tracking glasses in a physical driving setup and matches them with dashcam footage. While there are other eye tracking datasets for driving \citep{xia2018predicting}, they are often collected in simulated environments due to the difficulty of collecting and processing data in a naturalistic setting. The DR(eye)VE dataset is one of the most comprehensive datasets yet, and also includes human annotations indicating segments of time in which the driver is attentive towards driving-relevant stimuli or inattentive.

The conjunction of XAI and eyetracking has been rarely explored. \cite{schiller2020relevance} compared LRP heatmapping methods to eye gaze for facial expression recognition, finding some correlations between which parts of the face received the most attention and which parts the neural network found relevant. The results of this work indicate some promise in using visual LRP heatmaps as a means for human-machine teaming. Compared to the task of expression recognition, driving requires rapid task-dependent attention and motivates the creation of a closed-loop human-machine teaming system. As autonomous and semi-autonomous driving research increases, comparisons between humans and machines enable new technological development in this area.
\section{Methods}
\label{methods}
\begin{figure*}[h!]
  \includegraphics[width=0.7\textwidth]{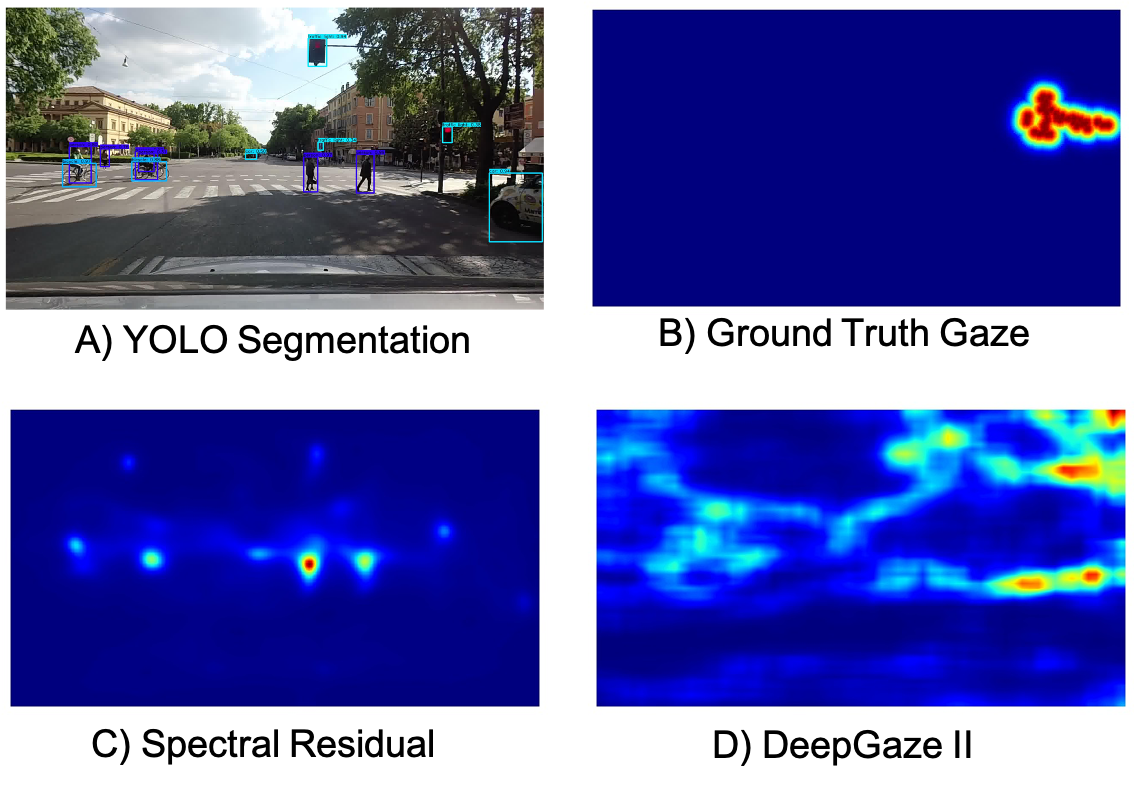}
  \caption{Heatmap generation techniques on a sample image. A) Original image with YOLO segmentation of objects such as pedestrians, bikes, and traffic lights. B) Ground truth gaze focus from DR(eye)VE dataset, which focuses on a traffic light. C) Saliency heatmap generated using spectral residual method, which seems to focus on pedestrians. D) Saliency heatmap generated using DeepGaze II method, which focuses on several areas including the traffic lights.}
  \label{heatmaps}
\end{figure*}
\begin{figure*}[h!]
  \includegraphics[width=0.7\textwidth]{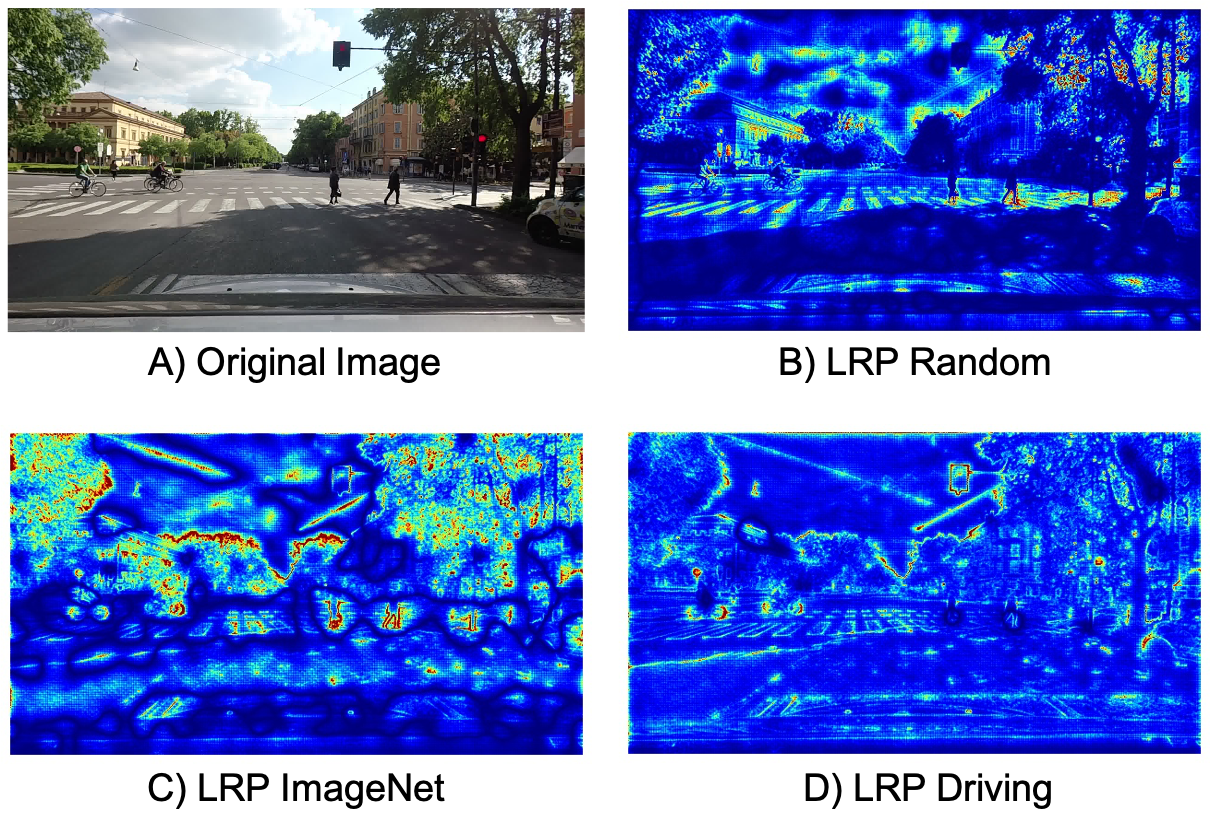}
  \caption{LRP heatmap generation. A) Original image. B) LRP heatmap of neural network with random weights, which seems to contain evenly distributed outlines. C) LRP heatmap of neural network pretrained on ImageNet, which outlines many object-related areas such as people and trees. D) LRP heatmap of neural network trained on driving output. Particular emphasis is found on the traffic light, which is not seen in C.}
  \label{lrpheatmaps}
\end{figure*}
\begin{figure}[h!]
  \includegraphics[width=0.47\textwidth]{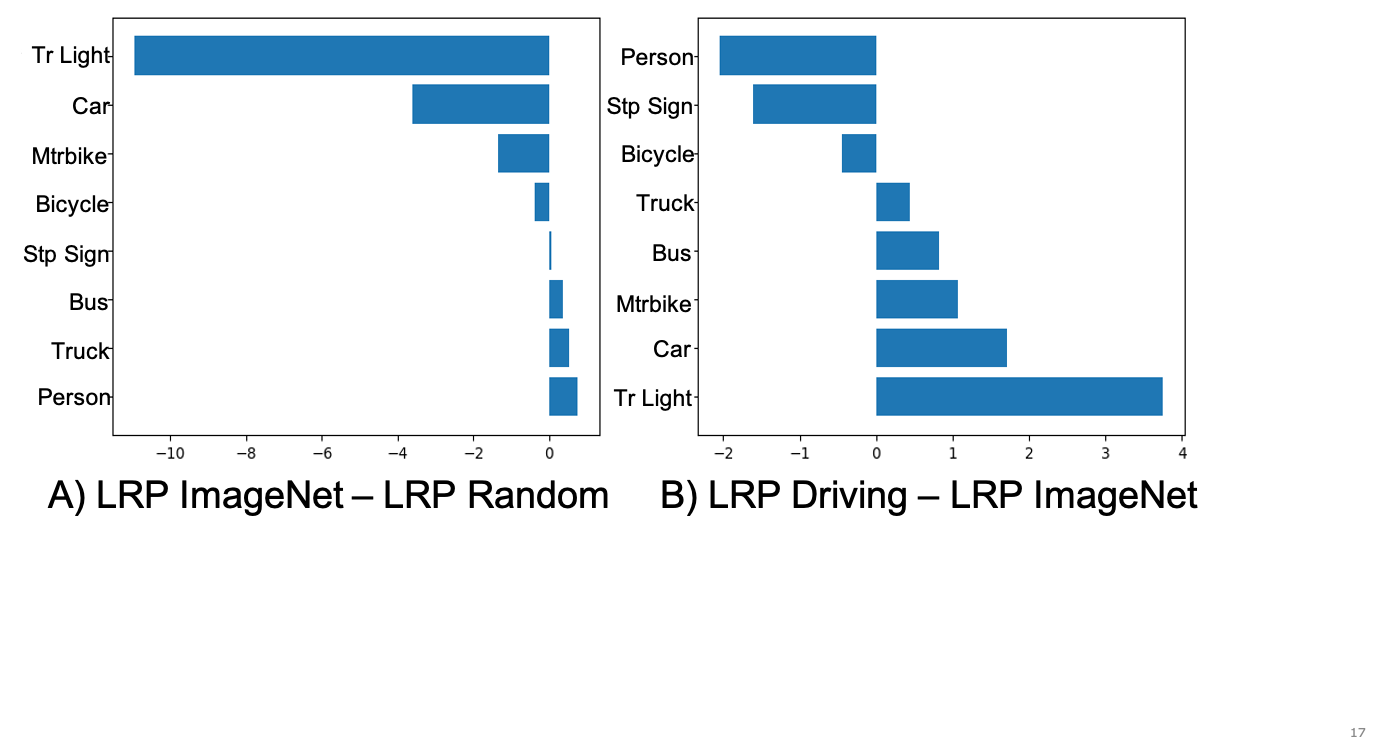}
  \vspace*{-10mm}
  \caption{Comparison of emphasis on traffic-related objects between heatmapping methods. X-axis contains differences in heatmap values within bounding boxes of objects, in units of e-7. A) Mean emphasis difference between LRP ImageNet and LRP Random. B) Mean emphasis difference between LRP Driving and LRP ImageNet. Larger emphasis on road-centered objects such as traffic lights and vehicles.}
  \label{objdifferences}
  \vspace*{-5mm}
\end{figure}
\begin{figure*}[h!]
  \includegraphics[width=0.7\textwidth]{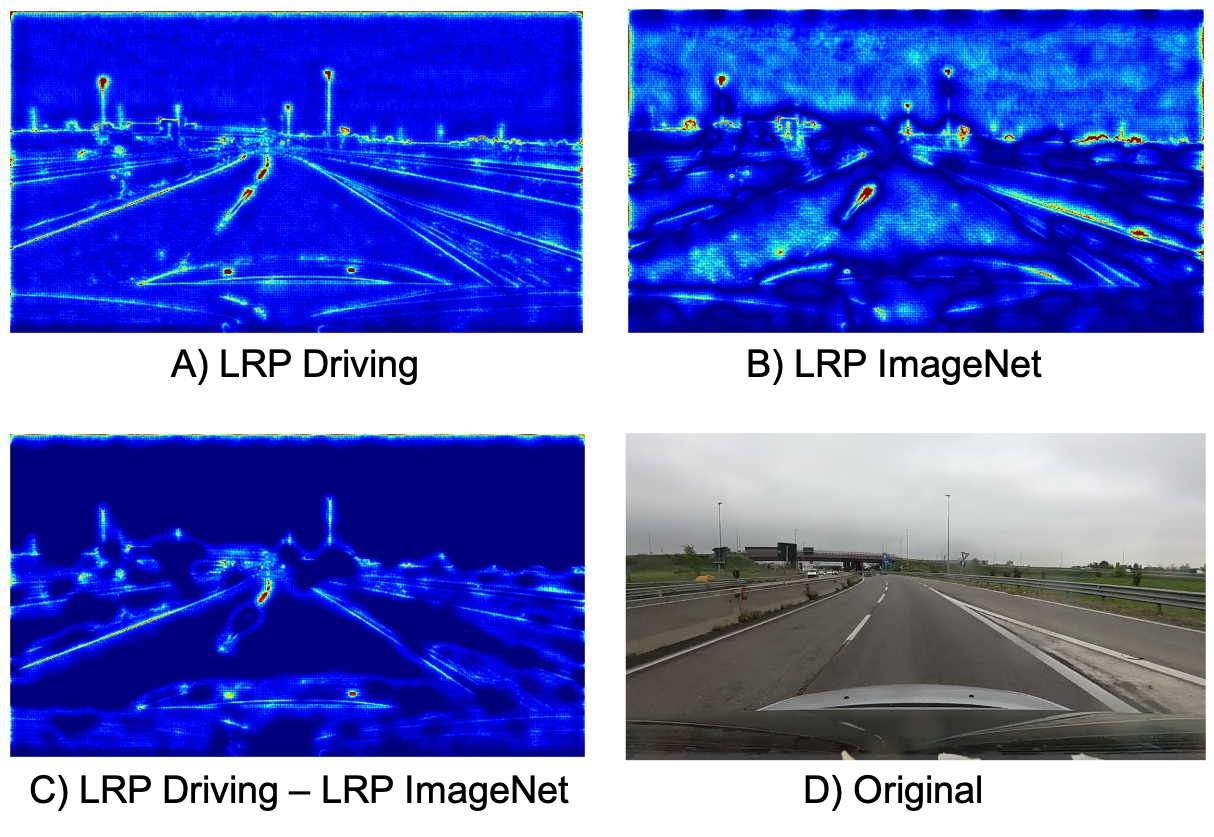}
  \caption{Example comparison between LRP Driving and LRP Imagenet on highway. A) Heatmap for LRP Driving. B) Heatmap for LRP ImageNet. C) Heatmap for clipped difference between A and B, showing emphasis on lane features. D) Original image for reference.}
  \label{lanelines}
\end{figure*}

The analysis described here centers around the comparison of saliency heatmaps generated by methods that vary along the scale of bottom-up and top-down processing. Each method is described below.
\subsection{Ground Truth Gaze}
For each dashcam frame in the DR(eye)VE dataset, there exists a corresponding frame indicating the gaze focus of the driver, integrated over a short segment of time before and after the frame. The frames are grayscale with a resolution of 1920x1080, where whiter values represent more salient areas. Each of the other heatmapping methods were compared to this ground truth measurement of gaze.
\subsection{Spectral Residual Saliency}
To form a baseline, a basic method of saliency detection known as the spectral residual \citep{hou2007saliency} was selected. The method is independent of features, categories, and prior knowledge of the image, relying on spatial frequency to find areas of interest. The implementation from the OpenCV library was used to generate greyscale saliency maps of the dashcam frames of the same dimension as ground truth gaze \citep{opencv_library}.
\subsection{DeepGaze II}
This work also sought to compare against the heatmaps generated by an existing module for effectively predicting human gaze. The DeepGaze II model was ideal for this, and their implementation was used to generate the saliency maps.
\subsection{LRP}
The main heatmaps of interest were those generated by LRP explanations of a deep neural network trained on the dashcam images as input. The default VGG16 network was used as a base model \citep{simonyan2014very}, removing the output layer and adding a layer of global average pooling, dropout layer of rate 0.2, and a dense layer with output size 2. The final network structure is described in Table \ref{nnlayers}.
\begin{table}
\centering
\begin{tabular}{ c c c }
Layer & Output Shape & Parameters \\
\hline
input & 1080x1920x3 & 0 \\  
vgg16 & 33x60x512 & 14714688 \\
global average pooling & 512 & 0 \\
dropout & 512 & 0 \\
dense & 2 & 1026 \\
\end{tabular}
\caption{Neural Network Layers}
\label{nnlayers}
\vspace*{-5mm}
\end{table}

80 percent of shuffled frames in the DR(eye)VE dataset were used for training, with the remaining frames used for testing. Using the iNNvestigate toolbox \citep{alber2019innvestigate}, LRP heatmaps of the trained neural networks running inference on each of the testing frames were generated, creating saliency heatmaps of the same format as the other methods. To examine the network at various stages of training, the heatmaps were generated on three versions of the neural network: 1) Completely randomized weights according to TensorFlow defaults, 2) The VGG16 portion containing weights pretrained on the ImageNet dataset \citep{ImageNet_cvpr09}, and 3) The same as 2) but with the non-VGG16 layers fine-tuned to predict the steering and velocity of the vehicle at each frame. The fine-tuning of layers was trained with mean squared error loss to convergence at 5 epochs, batch size 8, and learning rate 1e-3. For 3), the DR(eye)VE dataset did not contain ground truth values for steering and acceleration. These values were estimated using monocular visual odometry to find the camera yaw and translation from the current frame to the next frame. For this, the implementation from https://github.com/avisingh599/mono-vo was used, rescaling the yaw and translation values to the range of 0 to 1.

For further analysis on whether any specific objects in the frames received more attention in the LRP heatmaps, Yolov3 was used \citep{redmon2018yolov3}, implemented in TensorFlow 1.0 from the GitHub repository at https://github.com/YunYang1994/tensorflow-yolov3/. The network was pretrained to segment and identify objects from the COCO dataset \citep{lin2014microsoft}.

\subsection{Comparison Methods}
In initial examinations of the DR(eye)VE dataset, it was found that many frames were largely similar, with no notable events and gaze directed towards the center of view. The set of frames was narrowed to include those annotated as attentive and non-trivial. In addition, videos from nighttime runs detected objects poorly on the pretrained Yolov3 model. A further filter was applied such that only frames used in the testing set of the neural network were used. Using only frames that met these conditions, this work ultimately generated all heatmaps on a dataset of 56266 images.

Following the methods of \cite{schiller2020relevance}, cosine similarity and Spearman correlation was used to compare the heatmaps to ground truth gaze. To calculate cosine similarity, the heatmaps were flattened into vectors \textbf{A} and \textbf{B} and compared as
\begin{equation}
cos(\theta) = \frac{\mathbf{A}\cdot\mathbf{B}}{||\mathbf{A}|| ||\mathbf{B}||}
\end{equation}
Similarly, Spearman correlation was calculated as
\begin{equation}
r_s=\rho_{rg_X,rg_Y}=\frac{cov(rg_X,rg_Y)}{\sigma_{rg_X}\sigma_{rg_Y}}
\end{equation}
in which $rg_X$ and $rg_Y$ are the ranks of the raw values within each vector, $\rho$ is the correlation coefficient, $cov(rg_X,rg_Y)$ is the covariance of the rank variables, and $\sigma_{rg_X}$ and $\sigma_{rg_Y}$ are the standard deviations of the rank variables.

Prior to analysis, a positive correlation was predicted between the task-specificity of the heatmapping technique and its similarity with the heatmap of the driver's gaze. It was also predicted that the ratio of similarity scores between attentive and inattentive frames for each state would increase with task specificity. This reflects the bottom-up and top-down nature of eye movements, with inattentive states likely to be driven by unexpected saliency and attentive states driven by the task.

\section{Results}
\label{results}
\subsection{Bottom-Up to Top-Down Relationship in Similarity Scores}
The main hypothesis of this work was that there would be a positive correlation between the similarity scores and the task-specificity of the heatmap generation method. Calculating similarity scores to compare each heatmapping technique as shown in Tables \ref{cosinemedian}-\ref{spearmanmedian}, this main hypothesis was evaluated, calculating both the median scores within all frames, attentive frames, inattentive frames, and the ratio of attentive and inattentive. DeepGaze II had higher similarity than other methods in many conditions. As a prior state-of-the-art technique, this was not unexpected, especially as it was tailored specifically towards gaze prediction while the other methods were not. For cosine similarity, the hypotheses were supported in several ways. Within the LRP methods, the similarity score increased as the task specificity increased from Random to ImageNet to Driving. Spectral Residual was considered to be the least task-specific, and also had the lowest similarity; while LRP Driving was the most task-specific had the highest cosine similarity. Looking at the ratio of similarity scores between attentive and inattentive states for LRP methods, it was found that the more task-specific the heatmap, the higher the ratio between attentive and inattentive states. An inattentive person may look at salient objects unrelated to the task, which would be best predicted by methods such as DeepGaze II, whereas an attentive person may look at objects related to the task, which could be better predicted by XAI methods. All findings were significant with p $<$ .005. The significance between heatmapping methods was calculated using one-way ANOVA, and the significance of ratios of median values were found with the two-sided Mann-Whitney test of log values of attentive and inattentive.

Examining Spearman correlation, the effects are not as apparent. Among the heatmapping methods, it was noted that DeepGaze II generally outperformed the other methods in similarity. As will be shown in qualitative examples, a plausible reason for these observations can be found in the fact that LRP heatmaps consist of diffuse outlines, while the other heatmapping methods have more concentrated focal points. Because of this, the rank variables used in the Spearman metric may not capture the relevant portions of the LRP heatmaps as well as the cosine metric.

\begin{table}
\begin{tabular}{ c c c c c }
Method & All & Attentive & Inattentive & Ratio \\
\hline
Spectral Residual & 0.00624 & 0.00619 & 0.00649 & 0.95524\\  
Deep Gaze II & 0.01332 & 0.01319 & 0.01422 & 0.92707\\
LRP Random & 0.00914 & 0.00906 & 0.00958 & 0.94570 \\
LRP ImageNet & 0.01114 & 0.01109 & 0.01144 & 0.96951 \\
LRP Driving & 0.01330 & 0.01337 & 0.01297 & 1.03069 \\
\end{tabular}
\caption{Cosine Similarity, Median}
\label{cosinemedian}
\vspace*{-5mm}
\end{table}

\begin{table}
\begin{tabular}{ c c c c c }
Method & All & Attentive & Inattentive & Ratio \\
\hline
Spectral Residual & 1222.1 & 1226.1 & 1196.4 & 1.02486\\  
Deep Gaze II & 1577.7 & 1578.8 & 1570.5 & 1.00525\\
LRP Random & 1022.7 & 1029.9 & 969.37 & 1.062488 \\
LRP ImageNet & 801.629 & 804.465 & 788.097 & 1.02076 \\
LRP Driving & 857.030 & 858.234 & 848.420 & 1.01156 \\
\end{tabular}
\caption{Spearman Correlation, Median}
\label{spearmanmedian}
\vspace*{-5mm}
\end{table}

\subsection{Qualitative Examples}
Figure \ref{heatmaps} shows an example of heatmap generation of a frame in which the driver is stopped at an intersection. The ground truth gaze in \ref{heatmaps}B shows that the driver is currently looking at the traffic light on the right. The spectral residual heatmap in \ref{heatmaps}C shows emphasis on pedestrians, and the DeepGaze II heatmap in \ref{heatmaps}D shows emphasis in several areas including the traffic lights, pedestrians, and bikes. Figure \ref{lrpheatmaps} shows the heatmaps generated from LRP. Compared to the others, these are far more diffuse, unlike the focal nature of gaze. However, they are revealing in which stimuli are task-relevant. The LRP Random heatmap in \ref{lrpheatmaps}B shows general diffuse outlines, the LRP ImageNet heatmap in \ref{lrpheatmaps}C shows particular emphasis on objects such as the pedestrians and trees, and the LRP Driving heatmap in \ref{lrpheatmaps}D shows more focal emphasis on traffic-specific areas, such as the traffic light, which is almost undetectable in LRP ImageNet. LRP ImageNet heatmaps contained more noise in background areas, likely due to the extra untrained layers added to the pretrained VGG16 network. While LRP heatmaps alone may not be good predictors of gaze, they may reveal areas of interest that other methods cannot.

\subsection{Emergent Emphasis on Traffic-Relevant Objects}
The emphasis on the traffic light led to questions of whether any particular objects received more emphasis in one heatmapping technique over another. To compare the emphasis between any two techniques, the proportion of heatmap activity within the bounding box for each frame in the dataset and for each bounded object in that frame was determined. The difference of this value between the two techniques was computed and then averaged for each object over all frames. Figure \ref{objdifferences} shows the mean emphasis differences for traffic-related objects between the methods of LRP ImageNet and LRP Random in \ref{objdifferences}A, as well as LRP Driving and LRP ImageNet in \ref{objdifferences}B. Confirming observations, traffic lights were indeed more prominent in the LRP Driving heatmaps. In addition, the LRP Driving heatmaps tended to emphasize objects found on the road in plain view of the driver. While stop signs, bicycles, and pedestrians may affect driving output, they are often observed in the periphery, only affecting driving if seen within the vehicle's planned line of travel. Thus, a dataset with more corner case situations could potentially lead to different results, while the DR(eye)VE dataset captures typical driving situations. Also of note, many regular signs in the distance were falsely detected as stop signs, hence not being emphasized the same way that traffic lights were. Of note, the neural network and LRP heatmaps themselves had no notion of object bounding boxes or object labels, and still emphasized them according to task. LRP heatmaps may lead to emergent discovery of task-relevant stimuli that are not be obvious to human evaluators.

\subsection{Subtractive Methods for Finding Emergent Task-Specific Features}
This work further looked at which aspects of the frames were emphasized more in the LRP Driving heatmaps compared to LRP ImageNet, not accounting for object segmentation. To do this, the heatmaps from each method were normalized, then the LRP Imagenet heatmap was subtracted from the LRP Driving heatmap. To improve visibility in the LRP heatmaps, which were originally in the range of 0-255, were clipped to the range 0-100. Figure \ref{lanelines} shows an example where this subtraction showed a particular emphasis on lane features such as painted lines and boundaries. Without any explicit identification of such features, they were detected with LRP. This provides a potential means for highlighting features to inattentive drivers, as the subtracted heatmap is sparse enough to pick targeted areas for visual prompting.

\section{Discussion}
\label{discussion}
These results show that XAI techniques, particularly LRP, may be valuable in building a human-machine teaming system for useful applications in machine operation such as driving. Specifically, the LRP heatmaps are able to identify task-relevant stimuli without explicit object segmentation and classification, allowing both low-level and object-based features to be emphasized. Combined with other techniques for estimating human eye gaze, the full spectrum of bottom-up salient features and top-down task-specific features can be addressed.

LRP heatmapping techniques alone do not predict human gaze as well as other methods, raising the question of why the LRP heatmaps could not also be trained with human gaze data. While this is an appropriate next step, we wished to examine the human and machine representations separately before determining how to merge the two. We expect that a combination of machine-based task training and human-driven eye data would lead towards the goal of human-machine teaming methods of driver assistance.

Driving alert systems for quickly orienting distracted drivers have been developed in the past, with emphasis on using multiple modalities including auditory, haptic, and visual cues \citep{maltz2004imperfect,walch2015autonomous}. Particularly, visual cues in heads-up displays can alert drivers without requiring them to switch gaze between the road and dashboard \citep{doshi2008novel}. With such alerting systems, it is crucial to avoid unnecessary distractions and false alarms. The techniques in this work help to filter sensory input by task relevance, enabling intelligent filtering of cues and alerts. Moreover, XAI and LRP techniques are agnostic to sensor type, and are able to explain neural networks processing any modality of information.

\section{Future Directions}
\label{futuredirections}
Examining the frames without temporal information was a necessary prerequisite for gauging the whether basic XAI methods such as LRP would be useful to explore. However, as human gaze involves integrating many focal points over time, we hope to extend our evaluation to temporal XAI methods. For example, rather than using the VGG16 architecture, which takes isolated frames as input, a recurrent network may be used instead. This may increase the similarity with human gaze, as a human may choose to look away from task-relevant stimuli once acknowledged and processed.
To continue the work in developing such a human-machine system, we hope to test such methods in closed-loop training of human drivers. This requires design considerations in how to generate LRP-based visual prompts in real time, while avoiding distracting or irrelevant information. This would involve a mental model of human state, which may include not only whether the human is attentive or not, but also which stimuli the human has already processed. As we have seen, the method is also dependent on the dataset, requiring training on corner cases. Advanced causal reasoning methods could potentially be combined with the XAI techniques for better handling of such corner cases.

\bibliographystyle{elsarticle-harv.bst}
\bibliography{references}

\end{document}